\documentclass[reprint,superscriptaddress,aps,prb]{revtex4-2}

\usepackage{graphicx}
\usepackage{hyperref}
\usepackage{changes}
\usepackage{makerobust}
\usepackage{xcolor}
\usepackage{amsmath}
\usepackage{amssymb}
\usepackage{mathtools}

\newcommand{\makeauthor}[2]{\newcommand{#1}[1]{{%
  \protect%
  \color{#2}{%
    \bfseries\begingroup\escapechar=-1\edef\x{\endgroup\string#1}\x:%
  } ##1}}%
  \MakeRobustCommand#1}
\makeauthor{\lk}{purple}
\makeauthor{\jp}{red}
\makeauthor{\fg}{blue}

\definechangesauthor[name=Matteo, color=brown]{1}
\definechangesauthor[name=Hendrik, color=blue]{2}

\newcommand{\npatch}{\emph{N}-patch}

\newcommand{\bvec}[1]{\boldsymbol #1}
\begin{document}


\title{The kagome Hubbard model from a functional renormalization group perspective}

\author{Jonas B.~Profe}
\email{profe@itp.uni-frankfurt.de}
\affiliation{Institut f{\"u}r Theorie der Statistischen Physik, RWTH Aachen University and
  JARA---Fundamentals of Future Information Technology, 52056 Aachen, Germany}
\affiliation{Institute for Theoretical Physics, Goethe University Frankfurt,
Max-von-Laue-Straße 1, D-60438 Frankfurt a.M., Germany}
\author{Lennart Klebl}
\affiliation{I. Institute for Theoretical Physics, Universit\"at Hamburg, Hamburg, Germany}
\author{Francesco Grandi}
\affiliation{Institut f{\"u}r Theorie der Statistischen Physik, RWTH Aachen University and
  JARA---Fundamentals of Future Information Technology, 52056 Aachen, Germany}
\author{Hendrik Hohmann}
\affiliation{Institut f\"ur Theoretische Physik und Astrophysik and W\"urzburg-Dresden Cluster of Excellence ct.qmat, Universit\"at W\"urzburg, 97074 W\"urzburg, Germany}
\author{Matteo Dürrnagel}
\affiliation{Institut f\"ur Theoretische Physik und Astrophysik and W\"urzburg-Dresden Cluster of Excellence ct.qmat, Universit\"at W\"urzburg, 97074 W\"urzburg, Germany}
\affiliation{Institute for Theoretical Physics, ETH Z\"{u}rich, 8093 Z\"{u}rich, Switzerland}
\author{Tilman Schwemmer}
\affiliation{Institut f\"ur Theoretische Physik und Astrophysik and W\"urzburg-Dresden Cluster of Excellence ct.qmat, Universit\"at W\"urzburg, 97074 W\"urzburg, Germany}
\author{Ronny Thomale}
\affiliation{Institut f\"ur Theoretische Physik und Astrophysik and W\"urzburg-Dresden Cluster of Excellence ct.qmat, Universit\"at W\"urzburg, 97074 W\"urzburg, Germany}
\author{Dante M.~Kennes}
\email{dante.kennes@mpsd.mpg.de}
\affiliation{Institut f{\"u}r Theorie der Statistischen Physik, RWTH Aachen University and
  JARA---Fundamentals of Future Information Technology, 52056 Aachen, Germany}
 \affiliation{Max Planck Institute for the Structure and Dynamics of Matter, Center for Free Electron Laser Science, 22761 Hamburg, Germany}

\date{\today}
\begin{abstract}
The recent discovery of a variety of intricate electronic order in kagome metals has sprouted significant theoretical and experimental interest. From an electronic perspective on the potential microscopic origin of these phases, the most basic model is given by a Hubbard model on the kagome lattice. 
We employ functional renormalization group (FRG) to analyze the kagome Hubbard model. Through our methodological refinement of FRG both within its \npatch{} and truncated unity formulation, we resolve previous discrepancies of different FRG approaches (Wang et al., 2013 vs. Kiesel et al., 2013), and analyze both the pure ($p$-type) and mixed ($m$-type) van Hove fillings of the kagome lattice.
We further study the RG flow into symmetry broken phases to identify the energetically preferred linear combination of the respective order parameter without any need for additional mean field analysis. Our findings suggest some consistency with recent experiments, and underline the richness of electronic phases already found in the kagome Hubbard model. We also provide a no-go theorem for a complex charge bond ordered phase in the single orbital kagome Hubbard model, suggesting that this model cannot capture aspects of orbital current phases.
\end{abstract}

\maketitle

\section{Introduction}
{The vanadium-based kagome metals $A$V$_3$Sb$_5$ ($A=$K, Cs, Rb) are the most studied class of layered kagome systems so far. A rich interplay between electronic correlations, electron-phonon interactions, geometric frustration and topology is believed to be pivotal in determining their properties \cite{Neupert2022_NatPhys,Jiang2022_NSR,Guguchia2023_npjQM,Hu2023_npjQM}. At $\sim 100$K \cite{Ortiz2020_PRL}, these compounds undergo a charge-order (CO) phase transition leading to a $2 \times 2$ in-plane reconstruction of the unit cell. The out-of-plane component of the CO might depend on the cooling rate and on the compound of the series analyzed \cite{Xiao2023_PRR}; it either is $\times 1$ \cite{Li2022_NatComm}, $\times 2$ \cite{Jiang2021_Nat_Mat,Li2021_PRX,Liang2021_PRX} or $\times 4$ \cite{Ortiz2021_PRX}. Muon-spin relaxation \cite{TRS_breaking,Khasanov2022_PRR}, magneto-optical Kerr measurements \cite{Xu2022_NatPhys} and polar Kerr results \cite{Hu2022_arXiv2} observe a broken time reversal symmetry (TRS) in this phase, with no signatures of magnetic ordering observed. Therefore, this class of kagome metals is a prime contender for realizing the long sought after spontaneous orbital currents reminiscent of the Haldane \cite{Haldane1988_PRL} and the Varma \cite{Varma1997_PRB} models the latter of which has previously been pursued in high-T$_c$ cuprates~\cite{CRPHYS_2021__22_S5_7_0,PhysRevLett.99.027005,PhysRevB.77.094511}. Recent high resolution polar Kerr studies, however, do not find any evidence of broken TRS in this state~\cite{Saykin2023_PRL}.

The controversy of early experimental findings seems to be the rule rather than the exception for this class of compounds. Indeed, several experiments find indications of a phase transition at $\sim 50$K within the CO domain: Some of them find the transition to a low-temperature nematic phase which might be a zero-momentum charge order \cite{Nie2022_Nat,Zheng2022_Nat,Wulferding2022_PRR}, while others observe a one-dimensional CO with $1 \times 4$ in-plane reconstruction \cite{Zhao2021_Nat,Li2022_PRB,Li2023_PRX}. Recent investigations do not find any transition around $50$K \cite{Liu2023_arXiv,Frachet2023_arXiv}, suggesting that the vanadium-based kagome metals might be at the ``tipping point'' of correlated orders~\cite{Guo2023_arXiv}, i.e., small perturbations such as strain or an external magnetic field might stabilize a state with \textit{slightly} higher energy than the ground state at pristine conditions.

The properties of the CO state appear to be intertwined with the superconducting phase found below $\sim 1$K, rendering it unconventional in nature. Although a recent experiment points towards conventional $s$-wave symmetry for the superconducting gap function of the kagome metals~\cite{Zhang2023_NanLett}, several theoretical studies based on the Hubbard model on the kagome lattice have suggested $d$- \cite{Wang2013_PRB}, $d + i d$- \cite{Yu2012_PRB} and $f$-wave symmetry \cite{Kiesel2013_PRL} for the gap function, which would follow earlier experiments obtaining evidence for an unconventional superconductor~\cite{Ortiz2020_PRL, yin2021superconductivity, Guguchia2023}. Furthermore, recent theoretical investigations~\cite{PhysRevB.108.144508} highlight that the distinction between $s$ and $d$ wave, based on their impurity response, is not as straight forward, hence undermining the argumentation in~\cite{Zhang2023_NanLett}.

Refined theoretical simulations might help to solve some of the experimental controversies. \textit{Ab initio} descriptions of this class of compounds represent a reasonable starting point to tackle the problem, and they can provide indications on the minimal models required to describe the salient properties of kagome metals~\cite{Fuchs_2020, Consiglio2022_PRB}. Even if density functional theory seems unable to distinguish the driving force for the CO, which might be electronic or phononic~\cite{Zhou2021_PRB,Wang2022_PRM,Wang2022_PRB,Ptok2022_PRB}, still it can provide important indications regarding the Fermiology of the compounds. In particular, two different kinds of van Hove singularities (VHS) \cite{vanHove1953_PR} are found in the proximity of the Fermi level, suggesting their relevant role for the stabilization of symmetry-broken phases~\cite{Hu2022_NatComm}. Despite the fact that more than one orbital per site might be required~\cite{Wu2021_PRL} and electron-phonon interaction might play an important role \cite{Ferrari_SK2022, Kang_2022,Subires_2023,PhysRevB.109.075130} to properly describe the kagome metals, many of the theoretical works consider the single orbital extended Hubbard model to describe the main physical properties of this class of materials. This assumption is based, inter alia, on the correct replication of the experimentally observed VHS and their distinct sublattice character, \textit{i.e.} pure (p-type) and mixed (m-type) sublattice occupation.
Further, even thou both kinds of VHS are close to the Fermi level in the real systems, several works considered just the p-type VHS, sometimes even neglecting the sublattice character of the states at the Fermi level~\cite{Denner2021_PRL,Park2021_PRB,Lin2021_PRB,Dong2023_PRB,Grandi2023_PRB}, but the role of the m-type VHS has been taken into account in more recent investigations \cite{Romer2022_PRB,Scammell2023_NatComm,Wu2024_PRB,Huang2024_PRB}.
In this work, we clarify the FRG phase diagram of the single orbital extended Hubbard model on the kagome lattice, confirming earlier SMFRG results~\cite{Wang2013_PRB}.
Given the relevance of both p- and m-type VHS for the physics of the kagome metals, we study the model at both fillings. We derive an analytical condition for the absence of TRS breaking \textit{at} the phase transition, in agreement with results based on point-group symmetry arguments derived for Ginzburg-Landau theories at a continuous phase transition~\cite{Birman1966_PRL}.

The paper is structured as follows. In Section~\ref{sec:model} and \ref{sec:methods} we introduce the employed model and give a short introduction to the methods used to examine this model (FRG), respectively. This is followed by Section~\ref{sec:phase} where we discuss the phase diagram predicted by truncated unity FRG (TUFRG) and relate it to prior results~\cite{Kiesel2013_PRL, Wang2013_PRB}. In Section~\ref{sec:ana} we proceed by an in depth analysis of three different phases found at the p-type VHS, providing a no-go theorem for a perturbatively generated TRS breaking charge order. We then recalculate the phase diagrams with \npatch{} FRG in Section~\ref{sec:npatch}, highlighting that the two methods yield compatible results when using a state-of-the-art implementation~\cite{profe_divERGe}. In Section~\ref{sec:conc} we conclude by summarizing the paper and giving an outlook on possible next steps.

\section{Model}\label{sec:model}
The simplest model possibly describing the rich orderings observed in the $A{\mathrm{V}}_{3}{\mathrm{Sb}}_{5}$ group of kagome metals is the kagome-Hubbard model. We visualize the lattice and Hamiltonian terms in Fig~\ref{fig:fig1} (d). The Hamiltonian is given as 
\begin{multline}
    H = t \sum_{\langle i,j\rangle,\sigma} c_{i,\sigma}^\dagger
    c_{j,\sigma}^{\phantom\dagger} + U\sum_{i,} n_{i,\uparrow} n_{i,\downarrow}
    \\
    {}+V\sum_{\langle i,j\rangle,\sigma\sigma'} n_{i,\sigma} n_{j,\sigma'} \,,
\end{multline}
with $c_{i,\sigma}^{(\dagger)}$ the fermionic annihilation/creation operator acting on site $i$  and spin $\sigma$, $t$ the nearest neighbor hopping amplitude which we chose as $t = 1$ and measure all energies in units of $t$ from now on. $U$ and $V$ are the on-site and nearest neighbor density-density interactions and $n_{i,\sigma}=c_{i,\sigma}^{\dagger}c_{i,\sigma}$ is the electron density operator on site $i$ and spin $\sigma$. 

In analog to experimental observations, two dissimilar VHS placed at the $M$ points appear in the band structure, see Fig.~\ref{fig:fig1} (a) and (c). The difference between the VHS is related to the presence of three sites in the unit cell in the kagome lattice, i.e., three distinct sublattices $A$, $B$ and $C$. When the hopping integral $t > 0$, the upper (lower) VHS is called p-type (m-type) because the Fermi surface at that filling is sublattice \emph{pure} (\emph{mixed}). This means that the Fermi surface has a single (mixed) sublattice character at each $M$-point, which prevents (favors) a nesting condition driven by the local Hubbard interaction, the so called sublattice interference mechanism}~\cite{Kiesel2012_PRB,Consiglio2022_PRB}.

\section{Methods}\label{sec:methods}
In this work, we apply two different flavors of functional renormalization group (FRG)~\cite{metzner_functional_2012, Dupuis_2021}. 
FRG is based on integrating flow equations starting from a solvable theory to the full solution of the problem. These equations are derived by introducing a cutoff function $R(\Lambda)$ in the single-particle propagator. At the starting point $\Lambda=\Lambda_0$ (here $\Lambda_0=\infty$), the action is rendered solvable. From this starting point, we successively integrate out the hierarchy of flow-equations until the cutoff is removed, thus resulting in the exact solution. For general models, we have to solve an infinite set of coupled differential equations. As a consequence we have to employ approximations that make the equations numerically tractable. In this paper, we utilize the sharp cutoff $R(\Lambda) = \Theta(|\Lambda|-\nu)$ allowing for an efficient implementation of the numerically most demanding parts of the flow.

Due to the truncation of the hierarchy, the flow has to be stopped once a coupling becomes too large --- if this divergence happens in a non-polynomial fashion, it signalizes a divergence of a susceptibility and thereby a phase transition. By analysing the interaction at the divergence, we extract the leading order parameter and predict the expected ordered phase. Here the channels signalize different orderings, each associated with a different fermionic bilinear. The particle-particle channel ($P$) is associated to cooper-pair bilinears, its divergence signalizes a superconducting transition. The crossed particle-hole channel ($C$) is proportional to a spin-operator bilinear, thus resulting in a magnetic order parameter. Lastly, the direct-particle hole channel ($D$) is proportional to a particle-number operator bilinear, indicating charge ordering, once the magnetic contribution is subtracted. 

The two types of FRG we consider in the following are both built upon a level-2 truncated formulation of FRG (vertex-flow FRG or ``RPA+''), i.e.,~we discard self-energy feedback and frequency dependencies, but keep the flow of the two particle interaction  $F^\Lambda_{1,2,3,4}(\bvec k_1, \bvec k_2;\bvec k_3)$. Thereby we arrive at the following set of equations for the three diagrammatic channels $\Phi^{x}$, $x\in\{P,C,D\}$,
\begin{widetext}
\begin{align}
\label{eq:mom_pp}
  \frac{d\Phi^{P,\Lambda}_{1,2,3,4}(\bvec{k}_1,\bvec{k}_2;\bvec{k}_3)}{d\Lambda} &{} =
    \frac{1}{2}F^{\Lambda}_{1,2,1',2'}(\bvec{k}_1,\bvec{k}_2;\bvec{k}')
        F^{\Lambda}_{3',4',3,4}(\bvec{k}',\bvec{k}_1+\bvec{k}_2-\bvec{k}';\bvec{k}_3)
    \dot{L}^{\Lambda}_{1',2',3',4'}(\bvec{k}', \bvec{k}_1+\bvec{k}_2-\bvec{k}')\;, \\
\label{eq:mom_ph}
\frac{d\Phi^{C,\Lambda}_{1,2,3,4}(\bvec{k}_1,\bvec{k}_2;\bvec{k}_3)}{d\Lambda} &{} =
    F^{\Lambda}_{1,4',3,1'}(\bvec{k}_1,\bvec{k}';\bvec{k}_3)
        F^{\Lambda}_{3',2,2',4}(\bvec{k}'+\bvec{k}_3-\bvec{k}_1,\bvec{k}_2;\bvec{k}')
    \dot{L}^{\Lambda}_{1',2',3',4'}(\bvec{k}',\bvec{k}'+\bvec{k}_3-\bvec{k}_1)\;, \\
\label{eq:mom_phc}
\frac{d\Phi^{D,\Lambda}_{1,2,3,4}(\bvec{k}_1,\bvec{k}_2;\bvec{k}_3)}{d\Lambda} &{}=
    -F^{\Lambda}_{1,4',1'4}(\bvec{k}_1,\bvec{k}'+\bvec{k}_2-\bvec{k}_3;\bvec{k}')
        F^{\Lambda}_{3',2,3,2'}(\bvec{k}',\bvec{k}_2;\bvec{k}_3)
    \dot{L}^{\Lambda}_{1',2',3',4'}(\bvec{k}',\bvec{k}'+\bvec{k}_2-\bvec{k}_3)\;,
\end{align}
where we defined the non-interacting two particle propagator as 
\begin{equation}
    \dot{L}_{1,2,3,4}^{\Lambda}(\bvec{k}_1,\bvec{k}_2,\bvec{k}_3,\bvec{k}_4) =
    \left[S_{1,3}^{\Lambda}(\bvec{k}_1)\,G_{2,4}^{\Lambda}(\bvec{k}_2) +
              G_{1,3}^{\Lambda}(\bvec{k}_1)\,S_{2,4}^{\Lambda}(\bvec{k}_2)\right]
    \delta_{\bvec{k}_1,\bvec{k}_3} \delta_{\bvec{k}_2,\bvec{k}_4}\;.
\end{equation}
\end{widetext}
The full vertex can then be restored by summing up the three channel contributions and the irreducible vertex contribution.
Differences between truncated unity FRG and \npatch{} FRG are detailed in Ref~\cite{Beyer_2022}. Importantly, the two different variants should give consistent results, which for the kagome~Hubbard model has not been the case~\cite{Wang2013_PRB,Kiesel2013_PRL}, a conundrum we  resolve in this paper.

\section{Phase diagram} \label{sec:phase}
\begin{figure*}[!thb]
    \centering
    \includegraphics[width=\textwidth]{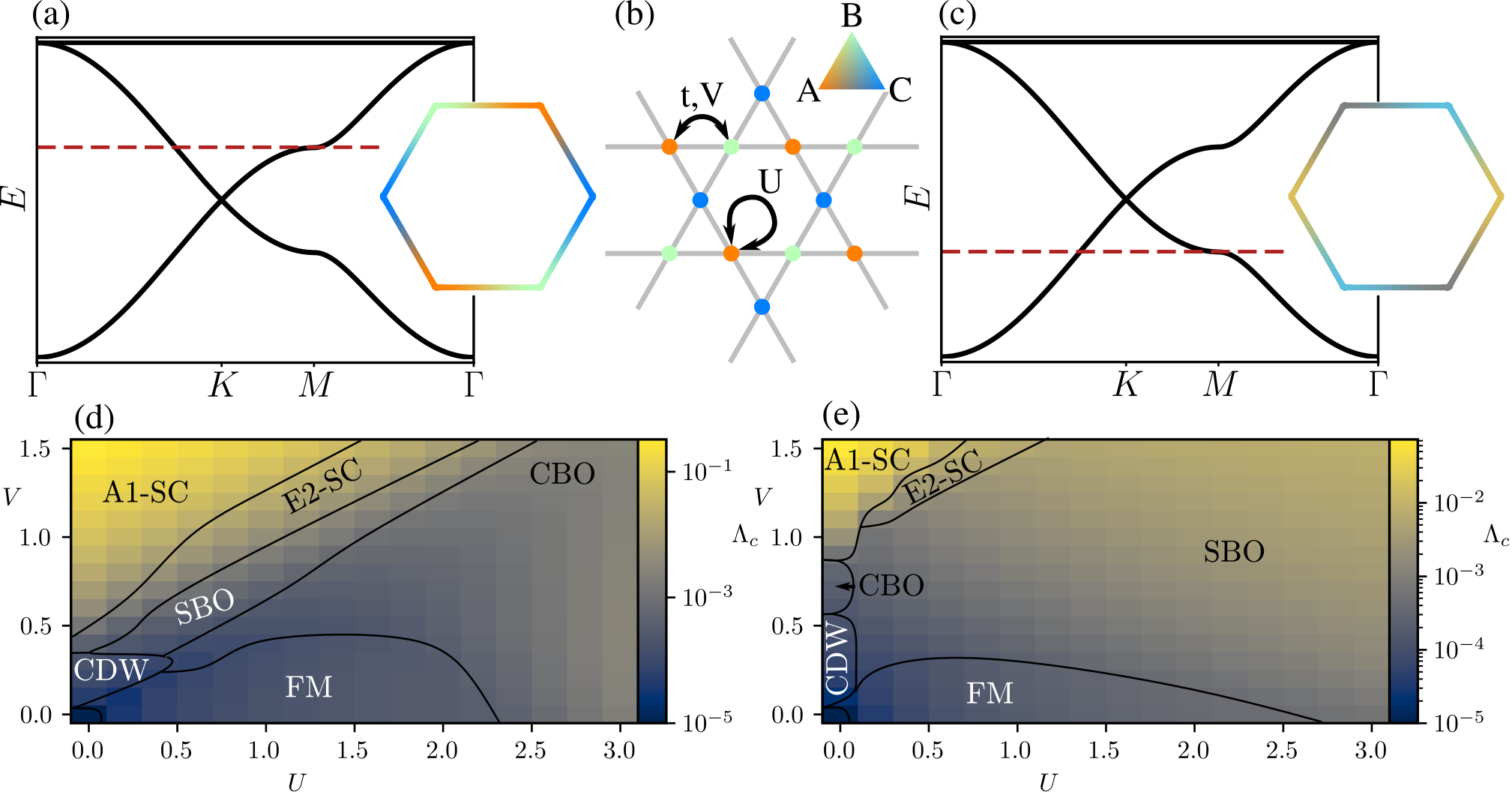}
    \caption{Bandstructure, orbital makeup of the FS statesat the $p$ and $m$-type VHS and schematic phase diagrams of the p and m-type filled kagome-Hubbard model. Fig (a) visualizes the bandstructure with the $p$-VHS filling marked in red (dashed) in red. The right panel admixture subfigure (a) depicts the orbital makeup at the Fermi level, with the colors encoding the admixture of the different sublattices according to Figure (b), that contains the lattice structure with all relevant parameters from the Hamiltonian. Fig. (c) is the same as Fig (a) but for the m-type VHS. Fig (d) shows the phase diagram in $U$-$V$ space at the p-type VHS. We find a Ferromagnetic region (FM), a charge density order (CDW), a charge bond order (CBO), a spin bond order (SBO), an $E_2$ superconductor (E2-SC) and an $A_1$ superconductor (A1-SC). The phase diagram at the $m$-type VHS in Fig (e) features the same phases as observed at $p$-type VHS.}
    \label{fig:fig1}
\end{figure*}

We begin by discussing the phase diagram at the two different van Hove types.
The prior discussed change in orbital makeup drastically changes the predicted phase diagram, see Fig.~\ref{fig:fig1} (d) and (e). In both we find the same phases, but their phase space volume is vastly different. We find a doubly degenerate $\bvec{q} = 0$ charge-density wave order at low $U$ and $V$, which will be discussed in detail in section~\ref{sec:ana}. For small nearest-neighbor interactions we find a large ferromagnetic region. This phase stems from the divergent density of states, which fuels a divergence of the particle-hole loop at low temperatures. In order to obtain this phase, the model's kinetics have to be finely resolved as otherwise the divergence is smeared out by limited momentum resolution. At large $V$, we find a superconducting order with 
$A_1$ symmetry (an s-wave) with uniform orbital weight on the three sublattices. At large $V$ the effective on-site interaction becomes attractive, strongly favoring double occupancy and a pair-formation to avoid the penalty of having neighboring electrons. Upon lowering $V$, we enter the $E_2$ superconducting state, which is discussed in much detail in Refs.~\cite{Wu2021_PRL, schwemmer2023pair} and briefly revisited in section \ref{sec:ana}. Notably, the on-site component of this order parameter does not need to vanish, as we analytically show in App.~\ref{app::onsiteE2}. We find that the order parameter weight is approximately evenly split between on-site and nearest neighbor bonds. The superconducting phases found agree with prior RPA studies~\cite{Romer2022_PRB}. The phase diagram we predict is in some sense complementary to the one found in variational monte carlo studies~\cite{Ferrari_SK2022} making comparisons difficult.

All phases mentioned before are located roughly within the same region of the phase diagram at the m and p type VHS - however the spin bond order and charge bond order are not. We name a phase spin or charge bond order if the leading ordering has non-zero weight on a bond and stems from either a crossed or direct particle-hole contribution. Both orders belong to the $A_1$ irreducible representation with transfer momentum $\bvec{q} = M$ (inducing a $2\times2$ enlargement of the unit-cell). They consist of on-site and bond components, mixing with different weights at different points in the phase diagram. At the p-type VHS, the charge bond-order makes up the largest portion of the phase diagram, while the SBO is driven by increased nearest neighbor interactions on top of the CBO. In contrast, at the m-type VHS, the SBO makes up most parts of the phase diagram, while the CBO is restricted to a very small region at small $U$ and intermediate $V$.

It should be noted, that these bond orders always mix with the respective on-site density wave. The ratio between bond and on-site order strongly depends on the chosen $U$ and $V$. To understand the origin of the phases, we examine the case $V = 0$ and unravel what drives the order, by subsequently turning off channels in the FRG calculation.
At the m-type VHS the dominant phase is an $M$-point SBO. The bond order character of the phase is induced by the interplay with the other diagrammatic channels: If we perform a flow for only the $C$-channel, we find an $M$-point spin-density wave roughly agreeing with the phase space of the spin-bond order. By inclusion of the $D$ channel, the critical scale changes by less than $1\%$ (at $U = 3t$), however now the observed order features bond contributions. This indicates that the spin order is primarily driven by RPA like diagrams, while the bond weights are generated by the higher harmonics induced from the feedback of the $P$- and $D$-channels. To understand why these higher harmonics are amplified, we need to consider the particle-hole loop at the $M$-point:
\begin{multline}
    \label{eq:ph_Mpnt}
    L_{o_1,o_3}^{b_1,b_3}(M) = \int d\bvec{k} \, e^{-i\bvec{k}(\bvec{B}_1-\bvec{B}_3)} \\
    (G_{o_1,o_3}(\bvec{k})G_{o_3+b_3,o_1+b_1}(\bvec{k}-M)+G\leftrightarrow G ).
\end{multline}
At low critical scales, the main contribution to the integral stems from the two nested Fermi surfaces (FSs) connected by $\bvec q=M$. Since at each point along the FS the weight is distributed between at least two orbitals, we obtain non-zero values for all components containing a suitable orbital combination---including the diagonal component $L_{o_1,o_1}^{0,0}$ and the on-site and bon mixing component $L_{o_1,o_1}^{0,b_i}$. Therefore, if the interaction contains a weak bond-order contribution arising from the inter-channel coupling it will get enhanced by the coupling present in the particle-hole loop.

In contrast, the charge bond order at the p-type VHS lacks a parental RPA-like phase, due to the sublattice interference mechanism preventing such an order~\cite{Kiesel2012_PRB}. If we only flow in the $D$ channel, we encounter no divergence. If we include the $C$ channel, we encounter a divergence in the $C$ channel at critical scales that are larger by an order of magnitude than for the full FRG flow. Therefore, the $P$-channel is a crucial ingredient in suppressing the ferromagnetic divergence. We can understand this again from the loop above. At the p-type VHS the Green's function has weight on at most two of the three orbitals at each point along the Fermi surface. Most importantly, in the high density regions, its weight is concentrated on a single orbital. This suppresses the on-site form factor components of the loop (but they are not zero). So no RPA like divergence at the $M$-point exists, since the high density regions mainly drive contributions in non-trivial form-factor sectors. Only once the bond components are generated from the $C$ and $P$ channel, we find the bond ordered phase. Note that this dependency on all channels renders the CBO less stable than the SBO. Further, this exemplifies the necessity of numerical studies beyond RPA approximations. 

\section{Analysis of orders}
\label{sec:ana}
In the following we give a detailed analysis of the charge orders as well as the $E_2$ superconducting order at the p-type VHS.

\subsection{Flowing into the symmetry broken phase}
After analyzing the leading instabilities from the TUFRG flow two different routes can be taken to extract more information about the form of the symmetry broken state. On the one hand, the FRG flow can be combined with a mean-field analysis of the effective model after the flow reached the critical scale $\Lambda_c$~\cite{PhysRevB.75.075110, o2023consistent}. Here, one needs to be careful in the construction of the divergence free part of the interactions.  On the other hand, one can include self-energies and allow for broken symmetries by insertion of a symmetry breaking perturbation at the initial scale~\cite{Markhof_2018, Gersch2006}. With this procedure one faces several challenges. 
First of all, the standard FRG flow equations are not mean-field exact~\cite{Salmhofer2004, Gersch2005, Gersch2006}. Only once the Katanin-substitution is introduced, the fulfillment of the Ward-identities is restored and the flow equations become mean-field exact~\cite{PhysRevB.70.115109}. This however is an issue, as we strongly rely on the sharp cutoff removing frequency integrals from our flow equations which is not possible anymore in the Katanin flow. We can therefore not assume \emph{a priori} that divergences can be removed by allowing broken symmetries. Past results for 1D models however do indicate that flowing into symmetry broken charge orders is possible without the inclusion of the Katanin substitution~\cite{Markhof_2018}.

Since the FRG flow is formulated in the grand canonical ensemble, enabling self-energy feedback implies a flowing particle number.
A solution that is frequently employed in self-consistent methods is to adapt the chemical potential after each flow step in order to force a given filling. In this pseudo-canonical picture, a counter term is introduced to the diagonal part of the self-energy, effectively altering the system in each step of the iteration.
While there are heuristic arguments why this is valid in self-consistent methods, it is known to break down in FRG away from particle-hole symmetry if the Katanin correction is not included~\cite{Rohe_2023}.
To circumvent possible issues arising from forcing an electron filling to the FRG flow, we perform a search where we vary the \emph{initial} chemical potential to arrive at the right filling value when the stopping condition is met.

\subsection{Charge orders}
\subsubsection{Charge density wave}

\begin{figure}[!htb]
    \centering
    \includegraphics[width = 0.95\linewidth]{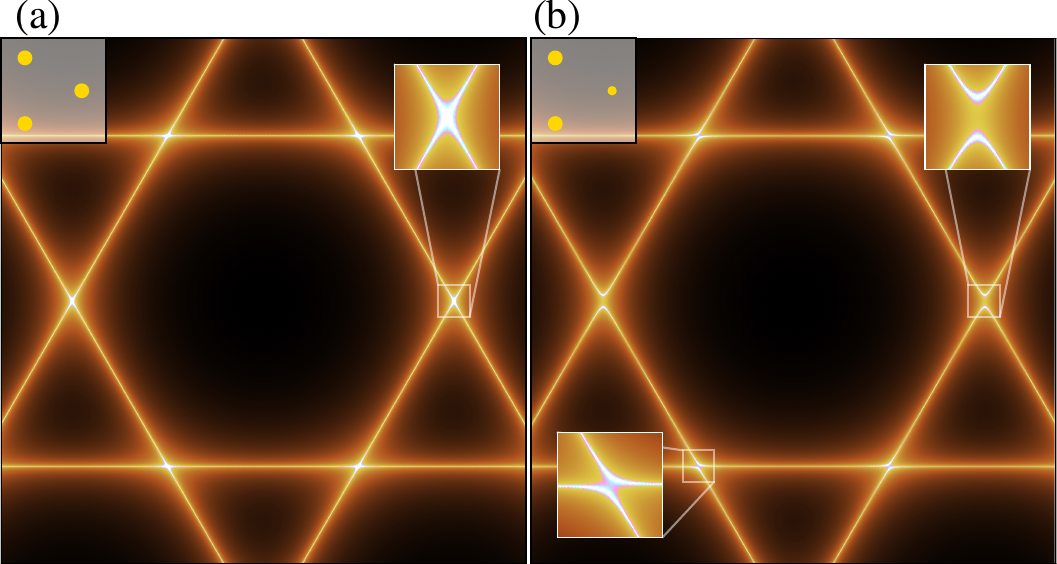}
    \caption{Spectral function $A(\bvec{k},\omega = 0)$ for the non-interacting (a) and symmetry broken state (b) respectively. In the upper left panels we visualize the occupation number in the two states on the three different sublattices. From a first FRG run, we found the CDW order to be in the $E_2$ irreducible representation. The two independent solutions are combined such that all $M$-points are fully gapped. }
    \label{fig:fig2}
\end{figure}

We first examine the simplest charge order: A charge density wave (orbital order) at $\bvec q=0$ that occurs for small $U$ and nonzero $V$.
The peculiar nature of the Kagome lattice allows for an on-site eigenvector that is in the $E_2$ irreducible representation. This order parameter hence acts as a site dependent chemical potential.
As the Hamiltonian has to be Hermitian, the two order parameters must not be superimposed in a complex fashion.
Since nontrivial superpositions of the two eigenvectors may occur nevertheless, we perform a flow into the charge ordered state and visualize the resulting spectral function in Fig.~\ref{fig:fig2}.

We observe a transition from a state in which all sites are equally populated to one in which one site is less populated. This charge reordering deforms the Fermi-surface to avoid the van-Hove singularities at the $M$-points, which effectively removes the divergence from the FRG flow. The gap-opening breaks the $C_{6v}$ symmetry down to a $C_{2v}$ symmetry, thus we find a Pomeranchuk like instability from electronic repulsion. Notably, if the flow did not open a gap at all VHS, we would expect a divergent susceptibility even in the symmetry broken phase.

\subsubsection{Charge bond order}
\begin{figure}[!hbt]
    \centering
    \includegraphics[width = 0.95\linewidth]{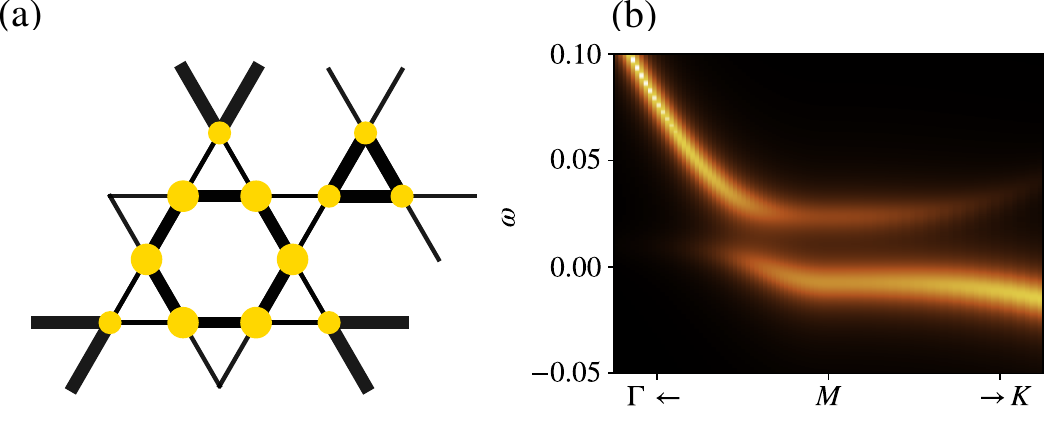}
    \caption{CBO flow into the symmetry broken phase. Left we show the predicted charge and hopping modulation in the symmetry broken phase, larger dots/connections indicate higher occupation/hopping. The spectral function around the M point on the right shows the formation of mini-bands at the Fermi-level. The slight asymmetry in the minibands stems from the non-fixing of the particle number during the FRG flow.}
    \label{fig:fig3}
\end{figure}

In experiments, a time reversal symmetry breaking charge bond order has been observed~\cite{TRS_breaking,Khasanov2022_PRR,Xu2022_NatPhys,Hu2022_arXiv2}.
Since this observation is under current debate, it is desirable to understand whether such a state can be facilitated by purely electronic effects in the single orbital model.
To answer this question with FRG, we set up a $2\times2$ supercell (with $12$ sites) that maps the $\bvec q=M$ charge order to the $\Gamma$ point. In this larger system, we repeat the
FRG flow with varying random initial symmetry breaking strength, initialized according to the form 
of the leading eigenvector of the instability in the symmetric phase. An exemplary result of an FRG flow into the symmetry broken phase is visualized in Fig.~\ref{fig:fig3}.

In all simulations, we observe a locking between the bond order and on-site components. The preferred configuration features higher occupation on the central hexagons and lower on the tips of the triangles. Furthermore, hopping between sites of the central hexagon is stronger than hopping out of the central hexagon (such a pattern was dubbed anti tri-hexagonal~\cite{Grandi2023_PRB}). The enlarged unit cell leads to the formation of mini-bands at the $M$-point, see Fig.~\ref{fig:fig3} (b). This gapping of the $M$-point again removes the divergence of the susceptibility from the flow allowing us to enter the symmetry broken phase. 

As we have a phase locking between charge bond and density order, no time-reversal symmetry 
breaking can emerge in this state (as otherwise the Hamiltonian becomes non-Hermitian). The locking stems from the particle-hole loop at the $M$-point,  
see Eq.~\ref{eq:ph_Mpnt}, which enters the linearized (charge channel) gap equation in the form-factor basis:
\begin{equation}
    \lambda \Delta_{o_1}^{b_1} = \Gamma_{o_1,o_1'}^{b_1,b_1'}L_{o_1',o_2}^{b_1',b_2}\Delta_{o_2}^{b_2},
\end{equation}
where $\lambda$ is the eigenvalue and $\Delta$ is the gap.
Here, we immediately observe that if we start with a pure bond order gap, the first matrix multiplication 
$L_{o_1',o_2}^{b_1',b_2}\Delta_{o_2}^{b_2}$ will result in a gap function mixing on-site and bond components, due to the finite weight of the loop in the bond-on-site mixing components mentioned above. Thus the real bond order instability is a general feature of the Spin-1/2 kagome Hubbard model - in other words this is a no-go theorem for a complex charge-density in the electronic single orbital kagome model. To find a complex order, we have to remove the on-site components of the interaction decoupling the bond from the density order. This is for example achieved by considering a spinless model. Alternatively, in more realistic models, this feature can be avoided by the orbital structure of the model under consideration.

\subsection{Superconducting orders}
\begin{figure}[!hbt]
    \centering
    \includegraphics[width = 0.95\linewidth]{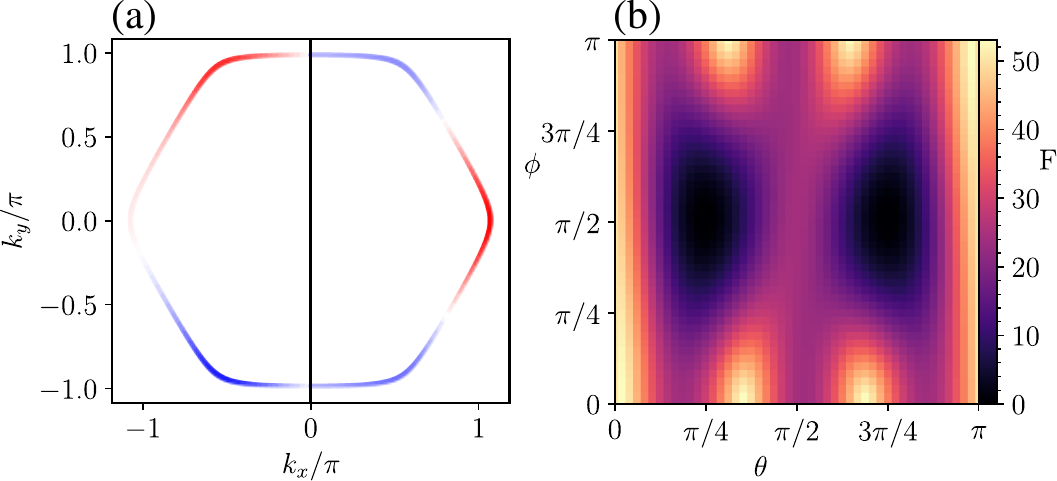}
    \caption{Chiral superconductivity. In (a) we show the leading eigenvectors from the linearized gap equation, each for one half of the Fermi-surface. We find two exactly degenerate eigenfuctions which obey a $C_2$ symmetry. The Gap equation is solved on the Fermi-surface. With the obtained linearized gap solutions, we calculate the free energy for different superpositions $\cos(\theta)\psi_1 + \sin(\theta)e^{i\phi}\psi_2$ resulting in (b). We see that chiral superpositions are preferred.}
    \label{fig:fig4}
\end{figure}

We found that flowing into the superconducting state is not possible without the Katanin substitution
as the divergence cannot be removed~\cite{Salmhofer2004}. Therefore, to analyse the preferred realization of the $E_2$ superconducting state we have to fall back to conventional methods. To find the energetically favored superposition we extract the gap functions from a linearized gap equation and feed them back into a single step self-consistent mean field equation, allowing us to track the free energy of every initial state, see Fig~\ref{fig:fig4}.

We observe minima in the free energy landscape at the $d\pm id$ superposition. I.e.~from this analysis the superconducting order is expected to be chiral~\cite{Nandkishore2012_NatPhys, schwemmer2023pair}. 

\section{\npatch{} results}\label{sec:npatch}
\begin{figure*}[!thb]
    \centering
    \includegraphics[width=\textwidth]{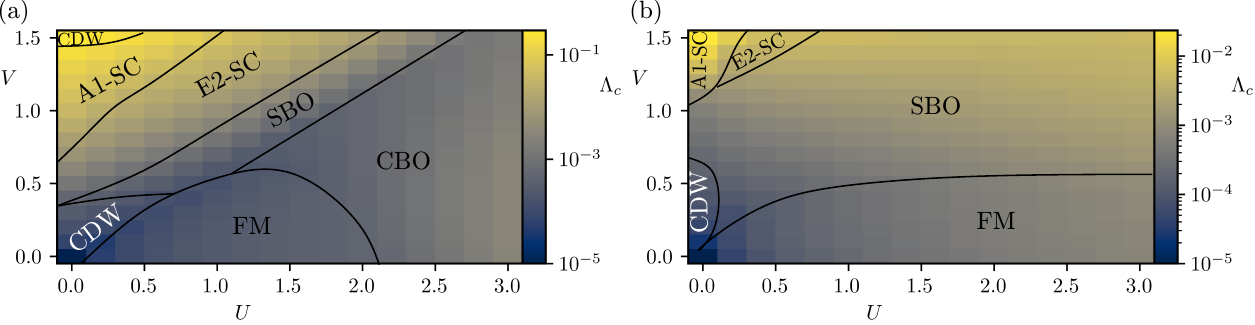}
    \caption{Schematic phase diagrams of the p and m-type filled kagome-Hubbard model calculated with \npatch. Fig a) shows the phase diagram in $U$-$V$ space at the p-type VHS. We find a Ferromagnetic region (FM), a charge density order (CDW), a charge bond order(CBO), a spin bond order (SBO), an $E_2$ superconductor (E2-SC) and an $A_1$ superconductor (A1-SC). Fig b) shows the phase diagram at the $m$-type VHS. The same phases as at the $p$-type VHS are observed, however the CBO is absent}
    \label{fig:fig5}
\end{figure*}
\npatch{} FRG is based on solving the flow equations on the Fermi-surface. 
This is motivated by a power counting argument~\cite{metzner_functional_2012}
which shows analytically, that only the contributions of the Fermi-surface of the two-particle 
vertex are not RG irrelevant. In practice, we represent the two-particle interaction as
\begin{equation}
    \Gamma_{o_1,o_2,o_3,o_4}(\bvec{k}_1,\bvec{k}_2,\bvec{k}_3),
\end{equation}
where the three momenta are restricted to the Fermi-surface. This however leads to a problem - in general $\bvec{k}_4$ is \emph{not} localized on the Fermi-surface. Thereby we implicitly break crossing symmetries, momentum conservation and all point group symmetries. Nonetheless, the application of patching RG has been fruitful~\cite{Kiesel_graphene,Honerkamp_patch}. 

To remedy these shortcomings partially, we can perform a resymmetrization of the vertex in each step of the flow. This procedure is well defined on a formal level since, as long as our patches are chosen according to the symmetries, all momenta the vertex explicitly depends on map correctly under symmetries. The fourth momentum enters the symmetry transformation exclusively as a phase prefactor, which we can calculate irrespective of the vertex parametrization:
\begin{multline}
    \Gamma_{o_1,o_2,o_3,o_4}(\bvec{k}_1,\bvec{k}_2,\bvec{k}_3) = \frac{1}{|\mathcal{G}|}\sum_{S\in \mathcal{G}} \\ S(\Gamma_{o_1,o_2,o_3,o_4}(\bvec{k}_1,\bvec{k}_2,\bvec{k}_3)) \,,
\end{multline}
where $S$ are the symmetry operations contained in the point-group of the lattice $\mathcal{G}$.
This procedure effectively removes the symmetry breaking and therefore allows us to observe correctly the degeneracies between the eigenvalues in two-dimensional irreducible representations of a point group.

Another subtlety necessitates the flow evaluation in orbital/sublattice space instead of band space. This requirement stems from missing gauge invariance of the two-particle vertex under the orbital to band transformation. In orbital space the matrix element interference is captured better (however still not completely as one finds analogously to Ref.~\cite{MEI_2019}). With these implementational advancements, we calculate the phase diagram, see Fig~\ref{fig:fig5}. At the p-type VHS the phase diagram agrees qualitatively with TUFRG and SMFRG, while at the m-type VHS, we observe two main differences: The ferromagnetic phase is enlarged and the small area of charge bond ordered phase is absent. 

As discussed above, the transition from the ferromagnetic to the spin bond order (and its on-site component) is visible in RPA, thus both phases emerge upon the fulfillment of the stoner criterion at either $\bvec{q_C} = 0$ or $\bvec{q_C} = M$. On the RPA level, the transition is observed in \npatch{} as well, only once the screening from the $P$-channel is incorporated, the FM is enhanced, while incorporating the $D$ channel leaves the phase diagram invariant. This can be understood as follows: The $D$-channel flow at $\bvec{q}_c = 0/M$, $\bvec{k}_1 = \bvec{k}_3 = 0$ has no momentum additions on the r.h.s. of Eq.~(3). Therefore it can be evaluated without approximations. On the other hand, the $P$-channel has a momentum addition---thus the vertex is evaluated at different parts of the Fermi-surface, leading to an overestimation of the screening at the $M$ point suppressing the transition to the spin-density wave. This highlights that for multi-site/multi-orbital  systems a pure \npatch{} approach is prone to approximation errors at larger scales at which contributions away from the Fermi-level are still relevant~\cite{MEI_2019}. A possible remedy is to switch from a truncated unity to an \npatch{} approach during the flow thereby merging the strengths of each of the methods.

To understand why our \npatch{} calculation does not agree with the earlier work by Kiesel \textit{et al.}~\cite{Kiesel2013_PRL}, we emphasize the difference in the approach taken. Here we stay in orbital space of  the full three site model, while the work by Kiesel \textit{et al.} works in a projected band space where the effective RG flow was reduced to the van Hove point carrying band. This turns out to be a too drastic approximation for some parts of the phase diagram, in particular when FRG seeks to identify instabilities in the particle-hole channel where the $\log^2$ divergence of the particle-particle channel is only overcome through intermediate coupling strength. With our methodological refinements which avoid the band projection altogether, \npatch{} FRG works more reliably and matches with the alternative truncated unity formulation.

\section{Conclusions}\label{sec:conc}
We examined the kagome Hubbard model with two different flavors of FRG, \npatch{} and truncated unity. Our results resolve the tension between earlier FRG results~\cite{Wang2013_PRB, Kiesel2013_PRL} by unifying the picture between TUFRG and \npatch{} highlighting the challenges mutli-orbital models pose in these approaches. 
Furthermore, we include static self-energy feedback and flow into the symmetry broken phases for both the charge density wave and the charge bond order. Here we found both to be a real superposition of the linearly independent order parameters. The CDW order leads to a deformed Fermi surface which gaps out the Van Hove singularities. The CBO realizes a $2 \times 2$ charge order that couples bond- with density sectors. Notably, this coupling is a general feature of the single orbital kagome-Hubbard model and provides the basis for the no-go theorem of a complex CBO. This highlights that either the single orbital kagome Hubbard model is not a suitable minimal model for the rich zoo of orderings observed in experiments, or that time reversal symmetry is not broken in the ground state of the system~\cite{Saykin2023_PRL, Guo2023_arXiv}. The lack of a flow to a charge ordered state and then to a superconducting phase in our simulations suggests that the single-orbital Hubbard model might not be the proper minimal model to describe the phenomenology of the vanadium-based kagome metals. 

In any case, it is of utmost importance to find a valid minimal model describing the physics at play in the kagome metals in order to unravel the puzzling experimental findings. Using models closer to the real materials and linking the FRG flow with \emph{ab-initio} simulations~\cite{profe_divERGe} seems to be the most promising route~.

\emph{Note added} Upon completing the manuscript we became aware of a recent publication studying  the $V=0$ line at the $m$-type van Hove singularity utilizing both SMFRG and variational monte carlo~\cite{PhysRevB.109.075127}, the results they obtained are in good agreement with our truncated unity FRG results.

\begin{acknowledgments}
We thank A. Fischer, M.~T.~Bunney, F.~Ferrari, J.~Beyer for fruitful discussions and R.~Valenti for carefully reading our manuscript.
This work was supported by the Excellence Initiative of the German federal and state governments, the Ministry of Innovation of North Rhine-Westphalia and the Deutsche Forschungsgemeinschaft (DFG, German Research Foundation). 
JBP, LK, FG, DMK acknowledge funding by the DFG under RTG 1995, within the Priority Program SPP 2244 ``2DMP'' --- 443273985.
LK and JBP greatfully acknowledge support from the DFG through FOR 5249 (QUAST, Project No. 449872909, TP5). LK, JBP and DMK gratefully acknowledge support from the DFG through SPP 2244 (Project No. 422707584).
DMK acknowledges support by the Max Planck-New York City Center for Nonequilibrium Quantum Phenomena. We acknowledge computational resources provided through the JARA Vergabegremium on the JARA Partition part of the supercomputer JURECA~\cite{JURECA} at Forschungszentrum Jülich.
HH, MD, TS and RT received funding from the Deutsche Forschungsgemeinschaft (DFG, German Research Foundation) through Project-ID 258499086 - SFB 1170 and through the W\"urzburg-Dresden Cluster of Excellence on Complexity and Topology in Quantum Matter -- \textit{ct.qmat} Project-ID 390858490 - EXC 2147.
\end{acknowledgments}
\bibliography{bib}
\appendix

\section{S-wave in E2}
\label{app::onsiteE2}
In the triangular and honeycomb lattice it is common knowledge that the $E_2$ irreducible representation does not allow for an on-site component of the order parameter. This follows directly from the requirement, that the mirror planes  have character $0$. In the kagome latttice we will prove in the following that this is not true, by explicitly constructing the system of equations and solving it.
We label the three sites a,b and c and the corresponding complex numbers $\alpha$, $\beta$ and $\gamma$. To calculate the character we have to apply the point group symmetry and calculate $\chi = \sum_i \bvec v_i^{*} \mathbf{O}\bvec v_i$ where i runs over the subspace dimension ($v_0 = v, v_1 = w$).
For the mirror planes we have $\chi = 0$. The mirrors exchange two sites, while the third maps onto itself. The $C_6$ rotation maps $a\rightarrow c,c\rightarrow b,b\rightarrow a$, The $C_3$ rotation maps $a\rightarrow b,b\rightarrow c,c\rightarrow a$. We directly observe that their inverse operations behave identically to the other rotation therefore the rotations give us instead of four conditions only two. The $C_2$ maps every site onto itself, and thus acts as an identity. We will assume normalized three component vectors, thus the $E/C_2$ condition is trivially fulfilled. An additional requirement is that the two vectors have to span a $2$D space.
The resulting set of equations read:
\begin{align*}
    0 &= v^*_a v_a + v^*_b v_c + v^*_c v_b + w^*_a w_a + w^*_b w_c + w^*_c w_b \\
    0 &= v^*_b v_b + v^*_a v_c + v^*_c v_a + w^*_b w_b + w^*_a w_c + w^*_c w_a \\
    0 &= v^*_c v_c + v^*_b v_a + v^*_a v_b + w^*_c w_c + w^*_b w_a + w^*_a w_b \\
    -1 &= v^*_a v_b + v^*_b v_c + v^*_c v_a + w^*_a w_b + w^*_b w_c + w^*_c w_a \\
    -1 &= v^*_a v_c + v^*_c v_b + v^*_b v_a + w^*_a w_c + w^*_c w_b + w^*_b w_a \\
    0 &= v^*_a w_a + v^*_b w_b + v^*_c w_c
\end{align*}

The equations are solved by $v_a = 0,  v_b =1/\sqrt{2}\cdot\phi, v_c = -v_b,
w_a = -2w_b,  w_b =1/\sqrt{6}\cdot\gamma, w_c = w_b$, where $\phi$ and $\gamma$ are global phases. As can be seen by insertion
\begin{align*}
    2|v_b|^2 &= 6|w_b|^2 \\
    |v_b|^2 &= 3|w_b|^2  \\
    |v_b|^2 &= 3|w_b|^2  \\
    -1 &= -|v_b|^2 + -3|w_b|^2 \\
    -1 &= -|v_b|^2 + -3|w_b|^2 \\
    0 &=  v_b^*w_b-v_b^*w_b. 
\end{align*}
The phases of $\bvec v$ and $\bvec w$ are free parameters.

\section{Existence of a ferromagnetic state}
\label{app:ferro}
In the following we will argue that the weak interaction ferromagnet, which was not seen in some recent studies, has to be present in the thermodynamic limit. Since we are interested in the weak-coupling limit, RPA arguments will suffice. In general we can rewrite the particle-hole loop at zero momentum transfer as
\begin{equation}
\begin{aligned}[b]
    L(0) &{}= \sum_{\omega , k}G(\omega,k)G(\omega,k) \\
         &{}= \sum_{k} n_f(\epsilon(k))(1-n_f(\epsilon(k))) \\
         &{}= \int d\epsilon\, \rho(\epsilon) n_f(\epsilon)(1-n_f(\epsilon)),
\end{aligned}
\end{equation}
where $\rho(\epsilon)$ is the density of states and $n_f(\epsilon)$ is the Fermi distribution. For a more general multi band model we need to project this loop onto the eigenvector corresponding to a FM, which has equal weight on all sites within the unit cell (we sum out all doubly occurring indices)
\begin{equation}
\begin{aligned}
    v_i L_{ij} v_j &{}= v_i G(\omega,k)_{ij}G(\omega,k)_{ji} v_j \\
                   &{}= \begin{multlined}[t]
                   \frac{1}{3}\sum_{k,\alpha,\beta} \frac{1}{i\omega-\epsilon(k,\alpha)}\frac{1}{i\omega-\epsilon(k,\beta)} \\
                   \sum_{i}U_{i,\alpha}(k)U_{i,\beta}(k)^*
                   \sum_{j}U_{j,\beta}(k)U_{j,\alpha}(k)^*
                   \end{multlined} \\
                   &{}= \frac{1}{3}\sum_{k,\alpha} \frac{1}{i\omega-\epsilon(k,\alpha)}\frac{1}{i\omega-\epsilon(k,\alpha)} \\
                   &{}= -\frac{1}{3T}\sum_{k,\alpha} n_f\big(\epsilon(k,\alpha)\big)\big[1-n_f\big(\epsilon(k,\alpha)\big)\big]
\end{aligned}
\end{equation}
Where $\epsilon(k,\alpha)$ is the dispersion of band $\alpha$ at momentum $k$ given as
\begin{align*}
    \epsilon(k,1) &= 2t \\
    \epsilon(k,2/3) &= t(-1\pm \sqrt{4A(k)-3}) \\ 
    \text{ with } A(k) &= \cos^2\left(\frac{k\cdot R_1}{2}\right)+\cos^2\left(\frac{k\cdot R_2}{2}\right)\\ &  +\cos^2\left(\frac{k\cdot (R_2-R_1)}{2}\right)
\end{align*}
with $R_1$ and $R_2$ being the basis vectors of the kagome lattice.
At the p-type VHS the flat band plays no role. Furthermore we can restrict the summation to an irreducible BZ wedge in which only one band crosses the Fermi level, here we pick for simplicity one in which band $2$ crosses. Thus we obtain
\begin{equation}
    v_i L_{ij} v_j = -\frac{4}{T}\sum_{k} n_f(\epsilon(k,2))(1-n_f(\epsilon(k,2))) 
\end{equation}
We can now rewrite this in terms of the DOS of the band as in Eq.~(\ref{eq:ph_Mpnt}):
\begin{equation}
    v_i L_{ij} v_j = -\frac{4}{3T}\int d\epsilon \rho(\epsilon) n_f(\epsilon)(1-n_f(\epsilon)).
\end{equation}
We have $n_f(\epsilon)(1-n_f(\epsilon)) = \delta_T(\epsilon)$ where $\delta_T(\epsilon)$ is a smeared out Dirac delta distribution, thus we get
\begin{equation}
    v_i L_{ij} v_j \approx -\frac{4}{T} \rho(0)
\end{equation}
Since we are at the van Hove singularity, the density of states is logarithmically diverging. This divergence is cut off by finite size effects explaining its absence in some earlier studies~\cite{Ferrari_SK2022}. In the TDL the ferromagnetic phase should be existent as long as no other phase gaps out the system beforehand.


\end{document}